\documentclass[aps,amsfonts,amsmath,prd,preprint,tightenlines,nofootinbib,showpacs]{revtex4}

\usepackage{mathrsfs,latexsym,bm}
\usepackage{pst-all}

\def\be{\begin{equation}}
\def\ee{\end{equation}}
\def\bea{\begin{eqnarray}}
\def\eea{\end{eqnarray}}
\def\bml{\begin{subequations}}
\def\blea{\bml\begin{eqnarray}}
\def\elea{\end{eqnarray}\end{subequations}}

\newcommand{\gb}{{\boldsymbol{g}}}

\newcommand{\ZZ}{{\mathbb Z}}

\begin{document}

\title{Averaged null energy condition in spacetimes with boundaries}

\author{Christopher J.\ Fewster}
\email{cjf3@york.ac.uk}
\affiliation{Department of Mathematics, University of York, Heslington, York, YO10~5DD, U.K.}

\author{Ken D.\ Olum}
\email{kdo@cosmos.phy.tufts.edu}
\affiliation{Institute of Cosmology, Department of Physics and Astronomy, Tufts University, Medford, MA  02155}

\author{Michael J.\ Pfenning}
\email{Michael.Pfenning@usma.edu}
\affiliation{Department of Physics, United States Military Academy, West Point, NY 10996}

\date{28 November 2006}

\begin{abstract}
The {\em Averaged Null Energy Condition} (ANEC) requires that the average
along a complete null geodesic of the projection of the stress-energy
tensor onto the geodesic tangent vector can never be negative. It is
sufficient to rule out many exotic phenomena in general relativity. 
Subject to certain conditions, we show that the ANEC can never be violated
by a quantized minimally coupled free scalar field along a complete null 
geodesic surrounded by a tubular neighborhood in which the geometry is
flat and whose intrinsic causal structure coincides with that induced
from the full spacetime.
In particular, the ANEC holds in flat space with boundaries, as in the Casimir
effect, for geodesics which stay a finite distance away from the boundary.
\end{abstract}


\maketitle

\section{Introduction}

General relativity alone gives no restriction on the geometry or topology
of spacetime.  Simply by solving Einstein's equation, $G_{ab}=
\kappa T_{ab}$, in reverse, one may easily determine the
stress-energy tensor $T_{ab}$ needed to sustain any desired
spacetime.  Thus the only possibility for restricting exotic phenomena
such as superluminal travel \cite{gr-qc/9805003}, traversable
wormholes \cite{Morris:1988tu}, or the creation of time machines
\cite{Hawking:1991nk,Tipler:1977eb} is to have {\em energy conditions}
that restrict $T_{ab}$.  Sufficient to rule out the exotic phenomena
above and to prove classical singularity theorems is the null energy
condition (NEC), which requires that $T_{ab}(x) V^a V^b\ge0$ for all
spacetime points $x$ and null vectors $V$.

Unfortunately, the NEC is violated when we introduce quantum fields;
an example is the original Casimir effect \cite{Casimir:1948dh}.  For
any $x$ between the (infinite) plates and any vector $V$ which is not
parallel to the plates, $\langle T_{ab}(x) V^a V^b\rangle$ will be
negative in the ground state.

A weaker condition 
 is the averaged null energy condition (ANEC), which
requires only that
\be\label{eqn:ANEC}
\int_\zeta dv\, T_{ab}k^a k^b\ge0\,,
\ee
where $\zeta(v)$ is a complete affinely parameterized null geodesic with tangent vector
$k$.  This condition is not expected to be violated by the Casimir plate system,
because geodesics parallel to the plates do not violate the NEC, while
any other geodesic eventually intersects the plate material, where it
picks up a positive contribution to the integral.
The ANEC is sufficient to rule out traversable wormholes and the
construction of time machines.  Variations of it can also be used to
prove singularity theorems
\cite{Galloway:1981kk,Roman:1986tp,Roman:1988vv} and to rule out
superluminal travel\footnote{The condition used in
\cite{gr-qc/9805003} is in fact the NEC, but one can see that is
sufficient to have the ANEC with the average over the path to be
traveled.} \cite{gr-qc/9805003}.

Various proofs of the ANEC for quantum matter are known. In
$n$-dimensional Minkowski space, the ANEC was proved for the scalar field
with arbitrary curvature coupling for a dense set of states in the Fock space
of the Minkowski vacuum \cite{Klinkhammer:1991ki}; namely, all finite linear 
combinations of finite particle-number states of bounded momentum.  
Folacci has posited a similar argument for electromagnetism in four
dimensions \cite{Folacci:1992xg}. 
One may also obtain the ANEC as a limiting case of the averaged weak energy
condition (AWEC)\footnote{The AWEC requires that
(\ref{eqn:ANEC}) holds with $\zeta$ replaced by a complete timelike geodesic.} [which
holds in Minkowski space as a consequence of Quantum Energy
Inequalities] \cite{Ford:1994bj} by considering the null geodesic as a
limit of timelike curves; the relevant class of states is not so clearly
defined here and quite strong conditions are required at
infinity to make this work (see \cite{gr-qc/0209036} for some comments
in this direction).
A more general and quite technical argument \cite{Wald:1991xn}, using
techniques of algebraic quantum field theory, establishes the
ANEC in four-dimensional Minkowski space for a restricted class of
Hadamard states of the minimally coupled scalar field. In particular,
this goes beyond the states contained in the usual Fock space. 

In two dimensions more can be said: in Minkowski space one may 
prove the ANEC for general (interacting) quantum fields with mass
\cite{Verch:1999nt}, or for unitary, positive energy conformal fields \cite{Fewster:2004nj};
precise classes of states for which these results hold are delineated in
each case. 
In general globally hyperbolic two-dimensional spacetimes, \cite{Wald:1991xn}
established the ANEC on complete achronal null geodesics for
the minimally coupled free scalar field. This holds for arbitrary
Hadamard states in the massless case, and a restricted class of
Hadamard states for nonzero mass; in neither case was there any
requirement that the states should be vectors in a Fock space
representation. The achronal condition on the geodesic may be dropped
\cite{Yurtsever:1994wc} at the expense of passing to a ``difference ANEC''. 

It is known that the ANEC may be violated in general curved spacetimes
(e.g., see \cite{gr-qc/9604008}) and in semi-classical quantum gravity
\cite{gr-qc/9602052}. Indeed, wormhole solutions have been found in
which the spacetime curvature leads to the ANEC violation that supports
the wormhole \cite{gr-qc/0501105,gr-qc/9701064,hep-th/0202068}, but
these involve Planck-scale structures, and thus should require a full
quantum gravitational treatment.  If one avoids physics at the Planck
scale, exotic phenomena can exist only with a very large separation of
scales \cite{gr-qc/9702026,gr-qc/9510071}.
The ANEC is also violated
for the ground state of the Klein-Gordon field in flat space with a
periodic identification in one spatial direction along any null geodesic
which winds around the compact direction.
We will return to this case, which is essentially
the Casimir system without the plates, in Sec.\ \ref{sec:example}. 

The above results all refer to globally hyperbolic spacetimes, which are
by definition boundaryless. Several attempts
\cite{gr-qc/0205134,hep-th/0307067,gr-qc/0407006,hep-th/0506136,hep-th/0507013}
have been made to violate the ANEC with specific systems of free fields
with boundaries, but these have been unsuccessful.
Ref. \cite{hep-th/0307067} conjectured that the ANEC is obeyed for all
geodesics passing outside a localized potential in flat space.

We will prove here a stronger version of the conjecture of
\cite{hep-th/0307067}. Namely,  we will show that distant boundaries or
potentials do not affect the ANEC, and that distant spacetime curvature
likewise has no effect if it does not change the causal structure near
the geodesic.  By ``distant'' here we mean simply that the geodesic must
not approach arbitrarily close to such places.  

The central idea is that the presence of distant boundaries can only be
detected if one can send a signal to the boundary and receive a signal
in return.  Otherwise one cannot determine whether one is making
measurements in spacetime with boundaries or merely observing an
unusual state on an unbounded spacetime.  But no signal can leave a
null geodesic in flat space and later return to it (neglecting, for the moment,
situations with nontrivial topology).  Thus for
measurements only on the geodesic, the boundaries cannot be observed,
and such a measurement must correspond to a possible measurement on
the geodesic in Minkowski space.  But the ANEC is obeyed in Minkowski space, so
we expect the ANEC to be obeyed even if there are boundaries.

At a more technical level, the key ingredient is the Quantum Null
Energy Inequality (QNEI) proved in \cite{gr-qc/0209036}, which places
bounds on weighted averages of the null-contracted stress-energy
tensor along {\em timelike} curves, and the covariance properties of
this bound (and other related Quantum Energy Inequalities (QEIs)) as
discussed in \cite{math-ph/0602042}. We refer the reader to
\cite{math-ph/0602042} and references therein for general background
on QEIs.

We will not be able to prove the ANEC without additional
assumptions (which bridge the gap between the averages considered in 
\cite{gr-qc/0209036} and the ANEC integral), but we will prove that there cannot be a
timelike-separated family of geodesics on which Eq.\ (\ref{eqn:ANEC})
converges uniformly to negative values.  To have some practical
effect, one needs not merely a single geodesic on which the ANEC is
violated but some finite region of violation.  Furthermore, one
expects that infinitesimally separated geodesics violate ANEC in the
same way, so that the convergence is uniform.  Thus we feel that
ruling out uniform ANEC violation is sufficient for all practical
purposes.

The paper is structured as follows. In Sec.\ \ref{sec:proof} we
rigorously state our premises and prove the conclusion.  In Sec.\
\ref{sec:variants} we give a variant argument in which we assume
uniform continuity of $T_{ab}$ and then can prove the ANEC on a single
geodesic.  After discussing the example of cylindrical spacetime in
Sec.\ \ref{sec:example}, we conclude in Sec.\ \ref{sec:discussion}.

\section{Main result}\label{sec:proof}

\subsection{Geometrical assumptions}\label{sec:geom}

Let $(N,\gb)$ be a spacetime composed of a four-dimensional smooth
manifold $N$ (possibly with boundary) equipped with a Lorentzian
metric $\gb$ and choices of orientation and time-orientation. We do not
assume that $(N,\gb)$ is globally hyperbolic, although we do assume that
strong causality holds on the interior of $N$ and that 
the spacetime supports a suitable quantum field theory, as we explain in
more detail below. 

Our aim is to prove the ANEC along any complete null geodesic $\zeta$ in
$N$ which is sufficiently isolated from the boundaries and regions of
curvature. More precisely, we assume that the Riemann tensor vanishes
everywhere in a simply-connected open  
neighborhood $N'$ of $\zeta$ that has no intersection with the boundary
of $N$, and further that
\be\label{eqn:causalitysame}
J^+(p, N)\cap N' =J^+(p, N') \qquad \text{for all $p\in N'$}\,,
\ee
i.e., regions outside $N'$ have no
effect on the causal structure inside $N'$. (The causal future $J^+$ can be
replaced, equivalently, by the causal past $J^-$ in (\ref{eqn:causalitysame}).) 

Given these assumptions, we can define coordinates $(t, x, y, z)$ on
$N'$ such that $\zeta$ may be affinely parameterized as
$\zeta(\lambda) = (\lambda,\lambda,0,0)$, and in which the metric
takes the Minkowski form $ds^2=dt^2-dx^2-dy^2-dz^2$ on $N'$. Without
loss of generality it may also be assumed that the tetrad
$(\partial/\partial t,\partial/\partial x,\partial/\partial
y,\partial/\partial z)$ is oriented and time-oriented, so our
coordinates define a (time-)orientation preserving isometry $\psi:N'\to M'$, 
where $M'$ is a region of Minkowski space $M$.

Finally, we assume that $N'$ is large enough to contain a tubular neighborhood of $\zeta$
of the form
\be 
N'_r = \{(t, x, y, z)\, |\, y^2+z^2+(x-t)^2 < r^2\}  \label{eqn:Nr'}
\ee
for some $r>0$. Thus $N'$ is not permitted to ``taper off'' at
infinity, thereby (for example) excluding situations where $\zeta$ is
asymptotic to the boundary of $N$ at large values of the affine
parameter, or where it passes through a sequence of apertures of
diminishing radius. 

Note that condition (\ref{eqn:causalitysame}) can only hold
if the null geodesic $\zeta$ is achronal. For suppose
there are points $p,q$ on $\zeta$
with $q\in I^+(p,N)$. Then $q$ is an interior point of $J^+(p,N)\cap
N'$, but a boundary point of $J^+(p,N')$, so these sets cannot be equal.

\subsection{Assumptions on the quantum field theory}\label{sec:qft}

We study the minimally coupled scalar field of mass $m\ge 0$ on
$(N,\gb)$. As we have not assumed that $(N,\gb)$ is globally hyperbolic, we must
make some further assumptions about the existence of a reasonable
quantization of the theory on this spacetime. The guiding principle,
motivated by the analysis of \cite{Brunetti:2001dx}, is
that the theory should coincide with the usual quantization on open
globally hyperbolic subsets\footnote{In
\cite{math-ph/0602042} these were referred to as c.e.g.h.s.\ regions.} of
the interior of $(N,\gb)$. 
Here, a subset $U$ of $(N,\gb)$ is
globally hyperbolic if strong causality holds on $U$ and $J^+(p,N)\cap
J^-(q,N)$ is a compact subset of $U$ for all $p,q\in U$ (see Sec. 6.6 of
\cite{Ha&El}). In particular, if $U$ is also open, it may be
regarded as a globally hyperbolic spacetime in its own right, but our
condition also ensures that the intrinsic causal structure of $(U,\gb|_U)$
coincides with that induced from $(N,\gb)$. Thus, for example,
`commutation at spacelike separation' has the same meaning in the two
spacetimes. As we have assumed that strong causality holds at each point of
the interior of $N$, every such point has an open globally
hyperbolic subset as a neighborhood. 

We also restrict attention to states which restrict to open globally
hyperbolic subsets of the interior of $N$ as Hadamard
states. In particular, these conditions are met when $(N,\gb)$ is
globally hyperbolic and we use the usual quantization, with the
states of interest required to be Hadamard on the
whole of $(N,\gb)$, but there are more general situations where they
hold (see~\cite{F&Pinprep} for more details and references). These conditions are
sufficient for the renormalized stress-energy tensor to be defined
according to the usual point-splitting prescription. For simplicity,
we will refer to our states as Hadamard, although they correspond to a
generalization of the usual concept. The conditions given here would, we
believe, be satisfied in any reasonable quantum field theory on $(N,\gb)$; we
do {\em not} claim that they are sufficient conditions to guarantee the
existence of such a theory. 

The geometrical conditions on $\zeta$ and its neighborhood $N'$ permit
us to identify a useful class of open globally hyperbolic subsets of $(N,\gb)$: let $p,q\in
N'$ and suppose
that the Minkowski space `double cone' $J^+(\psi(p),M)\cap J^-(\psi(q),M)$ is
contained within $M'$. Then the open set 
\be
D_{p,q}=I^+(p,N')\cap I^-(q,N'),
\ee
is a globally hyperbolic subset of $(N,\gb)$. 
To see this, first note that strong causality holds on $D_{p,q}$ because
it holds on the interior of $N$.
Now choose $r,s\in D_{p,q}$ and consider $K=J^+(r,N)\cap J^-(s,N)$. By
property (\ref{eqn:causalitysame}) [and the equivalent version for
the causal past] the intersection $K\cap N'$ is simply
$J^+(r,N')\cap J^-(s,N')$, which is a subset of $D_{p,q}$.\footnote{We
use the facts that $J^+(r,N')\subset I^+(p,N')$ and $J^-(s,N')\subset
I^-(q,N')$.} Thus $K$ is covered by the disjoint closed sets
$\overline{D}_{p,q}$ and $N\backslash N'$; since it is connected, it
must in fact be contained entirely in $\overline{D}_{p,q}$. Thus $K=J^+(r,N')\cap
J^-(s,N')$ and is therefore compact, because the corresponding Minkowski
double cone is. Hence $D_{p,q}$ is an open globally hyperbolic subset of
the interior of $(N,\gb)$. 

The main technical tool we will use to prove the ANEC is a QNEI established in
\cite{gr-qc/0209036} (Thrm III.1) which showed that the quantity
\be
\int_\gamma d\tau\,\langle T_{ab}k^ak^b\rangle_\omega(\gamma(\tau))g(\tau)^2
\ee
is bounded from below as $\omega$ varies over the class of Hadamard
states, where $\gamma$ is a smooth future-directed timelike curve parameterized by
proper time $\tau\in I$, for $I$ an open interval of the real
line,\footnote{In \cite{gr-qc/0209036} $I$ was taken to be the whole
line, but the difference is inessential.} $k$ is a smooth null vector field defined on a tubular
neighborhood of $\gamma$ and $g$ is any smooth, real-valued function
compactly supported in $I$. We must also require that $\gamma$ may be
enclosed in an open globally hyperbolic subset of $(N,\gb)$, because the result of
\cite{gr-qc/0209036} was proved in the globally hyperbolic setting. 

In general the lower bound given in  \cite{gr-qc/0209036} is exceedingly
difficult to calculate for a general worldline in a generic spacetime
(it depends on the two-point function of a reference Hadamard state).
However, the bound simplifies considerably if $\gamma$ may be enclosed
in an open globally hyperbolic subset $N'$ of $(N,\gb)$ which is isometric to
an open globally hyperbolic subset
$M'$ of Minkowski space in such a way that the (time-)orientation
induced on $M'$ by the isometry coincides with the usual
(time-)orientation of Minkowksi space. Under these circumstances the
QNEI bound reduces to the form it takes in Minkowski space (see Thrm.\ II.6 and 
Cor.\ II.4 of \cite{math-ph/0602042}, which establish the required
covariance property) and this has a simple closed form for the case
where $\gamma$ is geodesic and $k^a$ is covariantly constant near
$\gamma$. These bounds will be employed below for regions of the 
$D_{p,q}$ form mentioned above. It is also worth noting
that \cite{gr-qc/0209036} also showed that we cannot expect a quantum energy
inequality to hold for averaging along a null geodesic.

\subsection{The ANEC}

We may now state our main result, assuming that the spacetime $(N,\gb)$,
the complete null geodesic $\zeta$ and its neighborhood $N'$ obey the
geometrical conditions of Sect.~\ref{sec:geom}, and that $(N,\gb)$
admits a quantized real scalar field satisfying the
conditions of Sect.~\ref{sec:qft}.

Define $\Phi_0(\lambda,t)=(t+\lambda,\lambda,0,0)$, so that
$\zeta(\lambda)=\Phi_0(\lambda,0)$, and consider the ANEC integral,
\be\label{eqn:ANECintegral}
A(t) = \int_{-\infty}^\infty d\lambda\,T_\omega(\Phi_0(\lambda,t)) 
\ee
where $T_\omega(q) =\langle T_{ab}(q)k^a k^b \rangle_\omega$ is the
renormalized expectation value of the stress-energy tensor $T_{ab}$, 
in Hadamard state $\omega$, at the point $q$ projected onto the null 
tangent vector $k = \partial\Phi_0/\partial\lambda$. Our main result is
the following.

{\em Theorem II.1.} If $\omega$ is any Hadamard state
then it is impossible for the integral in
Eq.~(\ref{eqn:ANECintegral}) to converge uniformly to negative values
in any interval of $t$ containing $0$. That is, ANEC cannot be uniformly
violated near $\zeta$.

We remark that, because $T_\omega(q)$ is continuous in $q$, $A(t)$ is also
continuous in any interval of $t$ for which the integral in Eq.~(\ref{eqn:ANECintegral}) converges
uniformly. So an equivalent statement is that the integral cannot
converge uniformly in an interval of $t$ containing $0$ and have
$A(0)<0$. Before giving the proof of Theorem II.1, we state an immediate consequence.

{\em Corollary II.2.} If $\langle T_{ab}\rangle_\omega$ is 
stationary with respect to $t$ in $N'$ 
and the integral in Eq.~(\ref{eqn:ANECintegral}) converges for $t=0$, 
then $A(0)\ge 0$, i,e., the ANEC holds on $\zeta$ in state $\omega$. 

In order to prove Theorem II.1 we first introduce some families of timelike
curves. Given any velocity $v$ with corresponding
boost $\gamma =1/\sqrt{1-v^2}$, define a family of timelike geodesic
segments
\be
\Phi_v(\eta,\tau) = (\eta+\gamma\tau,\eta+\gamma v\tau,0,0)
\label{eqn:Phi_v_def}
\ee
parameterized by proper time $\tau \in (-\tau_0, \tau_0)$ and
labeled by $\eta \in(-\eta_0, \eta_0)$. 
The choice of $v$, $\tau_0$ and $\eta_0$ will be discussed below. Now
consider the set
\be
C_{v,\eta}=I^+(\Phi_v(\eta,-\tau_0),N')\cap I^-(\Phi_v(\eta,\tau_0),N')\,.
\ee
Providing that $\tau_0<r$, where $N_r'$ is of the form
Eq.~(\ref{eqn:Nr'}), it is easy to show that the Minkowski space boosted double cone
$J^+(\psi(\Phi_v(\eta,-\tau_0)),M)\cap J^-(\psi(\Phi_v(\eta,\tau_0)),
M)$ is contained within the region $\psi(N'_r)\subset M'$; thus
$C_{v,\eta}$ belongs to the family of open globally hyperbolic subsets of $(N,\gb)$ discussed in
Sec.\ \ref{sec:qft}. 
  
Accordingly, the discussion above entails that the QNEI of
\cite{gr-qc/0209036} may be applied to the segment $\tau\mapsto
\Phi_v(\eta,\tau)$ as if it were in Minkowski space. 
Let $f(a)$ be any smooth real-valued sampling function with 
support only for $|a|< 1$ and normalization
\be
\int f(a)^2 da = 1\,.
\ee
Then, from Eq.\ (3.10) of \cite{gr-qc/0209036} [applied to $g(\tau)=f(\tau/\tau_0)$] we have
\be 
\int_{-\infty}^\infty d\tau\, T_\omega(\Phi_v(\eta,\tau)) f(\tau/\tau_0)^2 \ge
- \frac{(v_a k^a)^2}{12\pi^2\tau_0^4}\int_{-\infty}^\infty d\tau
f''(\tau/\tau_0)^2
\ee
where the timelike tangent vector $v = \partial \Phi_v /\partial\tau$,
null vector $k = \partial \Phi_v /\partial\eta$,
and the prime denotes the derivative of the function 
with respect to its argument.\footnote{This bound was derived in
\cite{gr-qc/0209036} for the massless field, but also applies in the case $m>0$.}
Thus $(v_a k^a)^2= (1-v)/(1+v) < 1/\gamma^2$, and so 
\be\label{eqn:qwei}
\int_{-\infty}^\infty d\tau\, T_\omega(\Phi_v(\eta,\tau)) f(\tau/\tau_0)^2 \ge
-\frac{F}{12\pi^2\gamma^2\tau_0^3}
\ee
where
\be
F =\int f''(a)^2\,da
\ee
is a manifestly positive constant.  

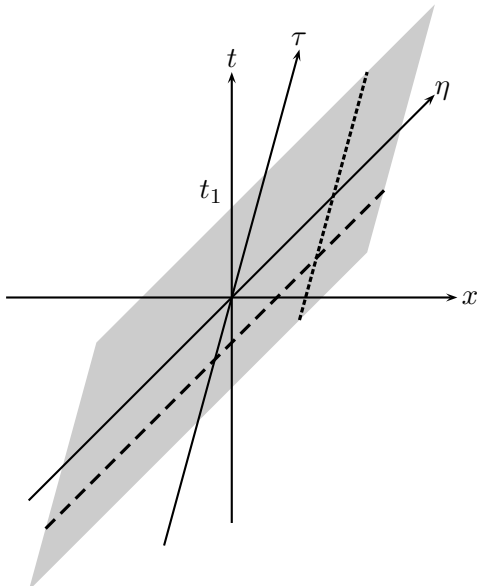
\begin{figure}
\begin{center}

\psset{xunit=3mm,yunit=3mm,runit=3mm}
\begin{pspicture}(-10,-11)(10,10)

\newgray{palegray}{.8}
\pscustom[fillstyle=solid,fillcolor=palegray,linestyle=none]{
   \psline{-}(-9,-13)(-6,-2)
   \psline{-}(-6,-2)(9,13)
   \psline{-}(9,13)(6,2)
   \psline{-}(6,2)(-9,-13)
   \closepath
}


\psline{->}(-10,0)(10,0) \rput[l](10.2,0){$x$} 
\psline{->}(0,-10)(0,10) \rput[b](0,10.2){$t$} 

\psline{->}(-9,-9)(9,9) \rput[l](9,9.2){$\eta$} 
\psline{->}(-3,-11)(3,11) \rput[b](3,11.2){$\tau$} 

\rput[rb](-0.4,4.2){$t_1$}

\psset{linewidth=1.2pt}
\psline[linestyle=dashed,dash=2pt 1pt]{-}(3,-1)(6,10) 
\psline[linestyle=dashed,dash=5pt 3pt]{-}(-8.25,-10.25)(6.75,4.75) 

\end{pspicture}
\end{center}

\caption{The parallelogram $\Phi_v(\eta,\tau)$, shown shaded, can be
considered as a set of timelike geodesics parameterized by $\tau$ with
$\eta$ fixed (short dashes), or of null geodesics
parameterized by $\eta$ with $\tau$ fixed (long dashes).}
\label{fig:parallelogram}

\end{figure}

The points $\Phi_v(\eta,\tau)$ with $-\tau_0 < \tau < \tau_0$ and
$-\eta_0 < \eta < \eta_0$ form a parallelogram, as shown in
Fig.\ \ref{fig:parallelogram}.  We can integrate over this parallelogram, 
\be\label{eqn:parallelogram1}
\int_{-\eta_0}^{\eta_0} d\eta\int_{-\tau_0}^{\tau_0}d\tau\,
T_\omega(\Phi_v(\eta,\tau))f(\tau/\tau_0)^2 \ge 
-\frac{\eta_0 F}{6\pi^2\gamma^2\tau_0^3}
\ee
By holding $\tau_0$ fixed and increasing $\gamma^2$ faster than
$\eta_0$, we can make the right hand side arbitrarily close to 0.  

We are now in a position to prove Theorem II.1. Suppose that the
integral in Eq.~(\ref{eqn:ANECintegral}) converges uniformly to negative values
in an interval of $t$ containing $0$. Thus, there are some positive numbers $A$ and $t_0$ (without loss of
generality, we take $t_0<r$) such that
\be
A(t) < - A \qquad\text{for all $t\in (-t_0, t_0)$.}
\ee

The parallelogram $\Phi_v(\eta,\tau)$ can also be considered as a
set of null geodesics parameterized by $\eta$ and labeled by
$\tau$.  Then $\Phi_v(\eta,\tau)=\Phi_0(\lambda, t)$ with the correspondence
\blea
t &=& (1-v)\gamma\tau\\
\lambda &=& \eta +\gamma v\tau
\elea
so our
parallelogram is $\Phi_0(\lambda, t)$ with $t\in (-t_1, t_1)$
and $\lambda \in (\lambda_-, \lambda_+)$ where
$t_1 = (1-v)\gamma\tau_0 =\tau_0/(\gamma (1+v))$ and $\lambda_\pm
=\pm\eta_0+ t v/(1-v)$.

Let us consider $\gamma$
sufficiently large that $t_1 < t_0$.  Then, since the integral in Eq.\
(\ref{eqn:ANECintegral}) converges uniformly, we can find a number
$\lambda_1$ such that
\be\label{eqn:parallelogram2}
\int_{-\eta_0}^{\eta_0} d\eta\, T_\omega(\Phi_v(\eta,\tau))
=\int_{\lambda_-}^{\lambda_+}  d\lambda\, T_\omega(\Phi_0(\lambda,t)) < -A/2
\ee
for all $t\in (-t_1,t_1)$, as long as
\bml\label{eqn:lambdapm}\bea
\lambda_+ &>& \lambda_1\\
\lambda_- &<&-\lambda_1\,.
\elea

By integrating Eq.\ (\ref{eqn:parallelogram2}), we find
\be\label{eqn:parallelogram3}
\int_{-\tau_0}^{\tau_0}d\tau\int_{-\eta_0}^{\eta_0} d\eta\,
T_\omega(\Phi_v(\eta,\tau))f(\tau/\tau_0)^2
< - \frac{A}{2}\int_{-\tau_0}^{\tau_0}d\tau
f(\tau/\tau_0)^2\ = - \frac{A\tau_0}{2}
\ee

Now choose any fixed, positive $\tau_0 < r$.  Let $\gamma$ grow
without bound and let $\eta_0 = \lambda_1+v\gamma\tau_0$ to satisfy
Eqs.\ (\ref{eqn:lambdapm}).  Then the
right hand side of Eq.\ (\ref{eqn:parallelogram1}) goes as
$\gamma^{-1}$, while that of Eq.\ (\ref{eqn:parallelogram3}) is
constant, so the two inequalities cannot simultaneously be satisfied in
this limit. This contradiction completes the proof of Theorem II.1. 

Another consequence of our result in this section is that if the state
$\omega$ is sufficiently regular that the integral in Eq.\ (\ref{eqn:ANECintegral}) converges
uniformly for all $t$ in a neighborhood of $t=0$ contained in $(-r,r)$, then we have
$A(t)\ge 0$ for all such $t$: this follows by applying Theorem II.1 to
the translates $\lambda\mapsto \Phi_0(\lambda,t)$ of $\zeta$, which also
satisfy the geometrical conditions of Sec.~\ref{sec:geom} (albeit with possibly smaller
tubular neighborhoods $N'_r$). 

\section{A Variant of the argument}\label{sec:variants}

In Sec.\ \ref{sec:proof} we showed that no Hadamard state can violate
the ANEC uniformly on a timelike family of geodesics, where violation of ANEC was
interpreted to mean that the ANEC integral converges to a 
negative value. It is also of interest to consider weaker
versions of ANEC (as in~\cite{Wald:1991xn}) in which convergence of the
integral is not required. The spacetime $(N,\gb)$, null geodesic $\zeta$
and neighborhood $N'$ are assumed to obey the conditions of
Secs.~\ref{sec:geom},~\ref{sec:qft}. Our result is the following. 

{\em Theorem III.1} Suppose that $T_\omega$ is uniformly continuous in
$N'$ for some Hadamard state $\omega$. Then 
\be\label{eqn:uniformcontinuityresult}
\liminf_{\lambda_0\to \infty} \int d\lambda\, T_\omega(\zeta(\lambda)) f(\lambda/\lambda_0)^2
\ge 0\,,
\ee
where $f$ is any smooth real-valued function supported in the interval $(-1,1)$. 

To prove this result, we approximate the null segment $\{\zeta(\lambda):|\lambda|\le
\lambda_0\}$, which contains the support of the integrand, by a timelike segment
$\Phi_v(0,\tau)$ [as defined in Eq.\ (\ref{eqn:Phi_v_def})]
with velocity $v$ and boost $\gamma$, and the
correspondence $\tau =\lambda/\gamma$. The velocity should be large
enough that $\lambda_0/\gamma<r$ in order to apply the Minkowski space
QNEI along the timelike curve. The distance between
corresponding points is
\be
|\zeta(\lambda)-\Phi_v(0,\tau)|=\gamma(1-v)|\tau| <|\tau|/\gamma
<\lambda_0/\gamma^2
\ee
and thus by uniform continuity,
\be\label{eqn:uniformcontinuity}
|T_\omega(\zeta(\lambda))-T_\omega(\Phi_v(0,\tau))|<C\lambda_0/\gamma^2\,,
\ee
where $C$ is a positive constant.  Integration gives
\be
\int d\lambda\, T_\omega(\zeta(\lambda)) f(\lambda/\lambda_0)^2
- \int d\lambda\, T_\omega(\Phi_v(0,\tau)) f(\tau/\tau_0)^2 > -2C\lambda_0^2/\gamma^2\,,
\ee
with $\tau =\lambda/\gamma$.  Changing variables in the second integral
and using Eq.\ (\ref{eqn:qwei}) gives
\be\label{eqn:doublebound}
\int d\lambda\, T_\omega(\zeta(\lambda)) f(\lambda/\lambda_0)^2
> -\frac{F}{12\pi^2\gamma \tau_0^3} -\frac{2C\lambda_0^2}{\gamma^2}
= -\frac{F\gamma^2}{12\pi^2\lambda_0^3} -\frac{2C\lambda_0^2}{\gamma^2}\,.
\ee

Now let $\lambda_0$ grow without bound, and let $v$ change so that $\gamma
=\lambda_0^{5/4}$ (thus $\lambda_0/\gamma\to 0$). The right hand side of Eq.\ (\ref{eqn:doublebound})
goes to zero, and we conclude that Eq.\
(\ref{eqn:uniformcontinuityresult}) holds, 
which is the form in which the ANEC was formulated
in~\cite{Wald:1991xn}. This completes the proof of Theorem III.1. 

A number of other hypotheses also lead to the conclusion Eq.\
(\ref{eqn:uniformcontinuityresult}). Noting that the difference between the points in Eq.\
(\ref{eqn:uniformcontinuity}) is purely in the $x$ direction, it is
sufficient to have $\partial T_\omega/\partial x$ bounded in place of
uniform continuity. Similarly, we could modify the proof so that the
timelike and null curves are related by the correspondence $\tau =\lambda/(v\gamma)$, in which case the
corresponding points differ only in $t$, and it is sufficient to have $\partial
T_\omega/\partial t$ bounded.  In particular, we can prove Eq.\
(\ref{eqn:uniformcontinuityresult}) in this way for any stationary configuration of
fields. Note also that the conclusion of Theorem III.1 applies equally
to any other complete null geodesic contained in $N_r'$.

\section{Example: the cylinder spacetime}\label{sec:example}

One of the simplest examples in which the ANEC is known to be violated is the
cylinder spacetime obtained by periodically identifying Minkowski space
along the $z$-direction. The ground state of the massless,
minimally coupled scalar field on this (globally hyperbolic) 
spacetime has renormalized stress tensor
\be
\langle T_{ab}\rangle = \frac{\pi^2}{90L^4}{\rm diag}\,(-1,1,1,-3)
\ee
where $L$ is the periodicity length. It is readily verified that the (A)NEC is
violated along any null geodesic with a nonzero $z$-component of
velocity, i.e., for those which wind around the compact direction. On
the other hand, the (A)NEC is obeyed along complete null geodesics with zero
velocity in the $z$-direction. We wish to show how this follows from our
main results.

First, note that null geodesics which wind round the cylinder are not
achronal, and are therefore excluded by our hypotheses. So the violation
of the ANEC along these curves is not in contradiction with our results.
Now consider any complete null geodesic $\zeta$ with zero velocity in the
$z$-direction. Without loss of generality, we may assume that its
three-velocity is directed in the positive $x$-direction. We will
construct a neighborhood $N'$ of $\zeta$ that satisfies the
conditions set out in Sec.\ \ref{sec:geom}. 

Let $M'=\{(t,x,y,z)\in M: (t-x)^2+y^2+z^2<r^2\}$ for some $r<L/4$. Then
we define $N'=\alpha(M')$, where $\alpha:M\to N$ is the quotient mapping that
implements the periodic identification. Because $|z|<r<L/2$ in $M'$,
$(t,x,y,z)$ are good coordinates on $N'$; moreover, $N'$ is simply
connected and does not taper off at infinity. Since the Riemann tensor
vanishes, we now need only check that condition
(\ref{eqn:causalitysame}) holds. Let $\tau_n$ represent a translation
through $L$ along the $z$-axis in $M$ and let $p$ be an arbitrary point
of $M'$. Then 
\be
J^+(\alpha(p),N)\cap N' = \alpha(M'\cap \bigcup_{n\in\ZZ} J^+(\tau_n p,M))
\ee
while 
\be
J^+(\alpha(p),N') = \alpha(J^+(p,M'))=\alpha(M'\cap J^+(p,M))\,,
\ee
where the equality $J^+(p,M')=M'\cap J^+(p,M)$ holds because $M'$ is
convex. Condition (\ref{eqn:causalitysame}) will follow if we can show
that $J^+(\tau_n p,M)\cap M'\subset
J^+(p,M)$ for each $n\in\ZZ$. This could fail only if
there were a point $q\in M'$ which was spacelike separated from $p$, but
timelike separated (in $M$) from some image point $p'$ of $p$. But the
difference in the $z$-coordinate of $p$ and $q$ must be less than $L/2$
(as both lie in $M'$) while the difference in the $z$ coordinate of $q$
and $p'$ must be greater than $L/2$. Thus the displacement from
$p'$ to $q$ must be
`more spacelike' than that from $p$ to $q$, and we see that no such $q$ can exist. 
Thus condition (\ref{eqn:causalitysame})
holds, and our results do apply to $\zeta$; not only in the ground
state, but in all Hadamard states on this spacetime. Moreover, they
would continue to apply even if we perturbed the metric on $(N,\gb)$
outside $N'$, provided the new metric respects condition~(\ref{eqn:causalitysame}).

\section{Discussion}\label{sec:discussion}

We have given a proof of the ANEC for  null geodesics with suitable
Minkowskian tubular neighborhoods, subject to conditions
which we feel are reasonable and do not impact its practical
applicability.  Our result is complementary to that of Flanagan and
Wald \cite{gr-qc/9602052}, who found that the ANEC is enforced in 
semiclassical quantum gravity up to second order in perturbation theory
about Minkowski spacetime, albeit with a modicum of transverse
smearing on the order of a few Planck lengths.

The main import of the present work is that distant interactions
or boundary conditions apparently do not help to produce ANEC
violations.  For example, Casimir plates would not seem to be
useful to support a wormhole, as in \cite{Morris:1988tu}, because of
the need to get through the plate \cite{hep-th/0506136}.

This paper is the first step in a program of proving that
wormholes, time machines and other such interesting phenomena
are not allowed within the domain of validity of the semiclassical
Einstein equation.  (``Self-supporting'' systems in which the geometry
itself will induce the ANEC violation necessary to produce it are
known \cite{gr-qc/0501105, gr-qc/9701064,hep-th/0202068}.  However,
they have consistently been shown to occur only on the order of the
Planck scale where the semiclassical approach is itself suspect, and
we must await a full quantum theory of gravity.)  We have handled the
case of free fields in flat space with boundary conditions.  We hope
in future work to be able to address interacting fields and curved
spacetimes.

\section*{Acknowledgments}

The authors would like to thank Larry Ford and Noah Graham for helpful
discussions.  K.~D.~O. was supported in part by the National Science
Foundation under grant 0353314 and by grant RFP1-06-024 from The
Foundational Questions Institute (fqxi.org) . M.~J.~P. would like to
acknowledge a grant for support of this work from the US Army Research
Office through the USMA Photonics Research Center.


\end{document}